\documentclass[twocolumn,showpacs]{revtex4-1}
\usepackage{epsfig}
\begin{document}
\title{Short-term synaptic facilitation improves information retrieval in noisy neural networks}
\author{J. F. Mejias$^1$, B. Hernandez-Gomez$^2$ and J. J. Torres$^2$\\     
  {\small $^1$ Department of Physics and Centre for Neural Dynamics,
University of Ottawa, K1N 6N5 Ottawa, Canada.}\\
{\small  $^2$ Department of Electromagnetism and Physics of the Matter,
University of Granada, 18071 Granada, Spain.
}}
\begin{abstract}
\begin{center}
\vspace*{0.5cm}
{\bf Abstract}
\end{center}
{Short-term synaptic depression and facilitation have been found to greatly
influence the performance of autoassociative neural networks. However, only
partial results, focused for instance on the computation of the maximum
storage capacity at zero temperature, have been obtained to date. In this
work, we extended the study of the effect of these synaptic mechanisms on
autoassociative neural networks to more realistic and general conditions,
including the presence of noise in the system. In particular, we characterized
the behavior of the system by means of its phase diagrams, and we concluded
that synaptic facilitation significantly enlarges the region of good retrieval
performance of the network. We also found that networks with facilitating
synapses may have critical temperatures substantially higher than those of
standard autoassociative networks, thus allowing neural networks to perform
better under high-noise conditions.}
\end{abstract}
\maketitle

\section{Introduction}
In the last years, there has been an increasing interest in the study of networked physical systems of excitable elements with time-varying connections between them. Such systems are often studied using the statistical mechanics formalism and techniques, and include many different physical, biological and social systems, such that ionic diffusion in magnetic materials \cite{torres98}, inter-relations among metabolic reactions within a cell or in food webs \cite{li10,larhlimi11}, logistic (transport and communication) networked systems \cite{kaufman98,tsekeris05}, and the friendship, social, professional and business relationships \cite{sun10,li11}. In many of these studies, the emergent behavior observed is a consequence of a non-trivial interplay between the excitability of units, the presence of intrinsic noise, and the nonlinear dynamics affecting the links among excitable units. 

A paradigmatical example of networked system relevant in this context is the so called autoassociative neural network (ANN). In the last years, there has been an attempt to revitalize the study of ANNs by including biologically realistic mechanisms among their features \cite{bibitchkov02,torresNC,matsumoto07,mejias09}. In particular, the consideration of activity-dependent mechanisms in their links or synapses, such as short-term depression (STD) and facilitation (STF) \cite{abbott97,zucker02} has revealed itself as a quite proficient topic. Examples of the rich and complex phenomenology found are the role of networks with depressing synapses as variable frequency oscillators \cite{vanrossum09}, the emergence of dynamical phases \cite{bibitchkov02,torresNC}, the associated instability of attractors consequence of the fatigue of the links \cite{fan07,cortes07}, the emergence, in some cases, of a chaotic itineracy among the attractors with optimal computational consequences \cite{defranciscis10}, the efficient detection of weak signals via stochastic multiresonances \cite{mejias11njp,mejias11plos}, the emergence of a critical dynamics in bistable systems driven by multiplicative colored noise \cite{mejias10plos}, or the study of storage capacity and retrieval properties of ANN with dynamic synapses \cite{bibitchkov02,torresCAPACITY,matsumoto07,mejias09,okada11}, to name a few. Most of these studies usually tackle the problem by considering Hopfield \cite{hopfield82} networks, which frequently allow to obtain information both analytically and numerically for the particular problem of interest. Quite often, to obtain such information one has to assume restricted conditions. For instance, to investigate the role of STD and STF on network dynamics, many studies have focused on networks with a finite number of stored patterns $P$ (so that the network load, defined as $\alpha \equiv P/N$ with $N$ being the number of neurons in the network, tends to zero in the thermodynamic limit $N \rightarrow \infty$) \cite{torresNC, cortes07}. Other studies consider, for instance, systems where the temperature parameter $T$ remains low (or even zero, which corresponds to deterministic dynamics) in order to evaluate maximal retrieval abilities of neural networks \cite{bibitchkov02,matsumoto07,mejias09}. However, results on the influence of STD and STF on neural networks in more general situations, such as with strong noise and nonzero network load, are still lacking. They could be particularly interesting in the case of STF, since it has been found to have positive effects in the critical storage capacity of ANNs \cite{mejias09}, in opposition to STD \cite{torresCAPACITY}. 

In this work, we study the effect of dynamic synapses, and in particular STD and STF, in the properties of ANN in general conditions of temperature and network load. To do this, we have numerically computed the phase diagrams of the system, and we have used a {\em na\"{i}ve} mean-field approach to interpret the results. We find that STF substantially improves the retrieval properties of the network, by increasing the area of the $\{ T,\alpha \}$ region of the phase diagram in which the system is able to efficiently recover stored activity patterns. This {\em memory phase} continues increasing its area even for conditions in which STF no longer improves the storing abilities of deterministic networks, suggesting that STF has a stronger impact in noisy neural networks than in deterministic ones. Networks with depressing synapses, on the other hand, have a smaller memory phase compared with standard Hopfield networks (that is, networks with no short-term synaptic plasticity). Finally, we particularize to the case of $\alpha \rightarrow 0$, and we demonstrate that the critical temperature $T_c$ increases with the level of synaptic facilitation. Although the absolute strength of synaptic couplings is upper-bounded as in classical ANN models, the particular value of $T_c$ may be arbitrarily high in our model, with the particular value depending on synaptic parameters. This suggests a key role of STF in situations in which previously stored memory has to be maintained in spite of noise, such as in working memory tasks \cite{compte09,mongillo08}.

\section{The model}

We consider a fully connected network of $N$ binary neurons whose states $s_i=0,1;\, \forall i=1,\ldots,N$ representing, respectively, silent or firing neurons, follow a probabilistic parallel, synchronous dynamics~\cite{perettoB}
\begin{equation}
P[s_i(t+1)=1]=\frac{1}{2}\left \{1+\tanh[2\beta (h_i({\bf s},t)-\theta_i)]\right \},
\label{prob}
\end{equation}
with $i=1\ldots N$, and where $h_i({\bf s},t)$ is the local field or the total input synaptic current to neuron $i$, namely
\begin{equation}
h_i({\bf s},t)= \sum_{j\neq i} \omega_{ij} x_j(t) u_j(t) s_j(t).
\label{lfp}
\end{equation}
We denote here $\beta=T^{-1}$ as the inverse of the temperature (or noise) parameter, in such a way that $\beta \rightarrow \infty$ implies deterministic dynamics. The quantity $\theta_i$ represents the neuron firing threshold of the neuron $i$, and the coefficients $\omega_{ij}$ are fixed synaptic conductances, determined by a slow learning process of $P$ patterns of activity. To model such learning process, we choose a hebbian prescription given by the standard covariance rule~\cite{tsodyksstorage88}
\begin{equation}
\omega_{ij}=\frac{1}{Nf(1-f)}\sum_{\mu=1}^{P} (\xi_i^{\mu}-f) (\xi_j^{\mu}-f),
\label{hebb}
\end{equation}
where $\left\lbrace \xi_i^{\mu}=0,1;\,i=1\ldots N\right\rbrace $ represents the $P$ stored random patterns with mean activity $\langle \xi^\mu_i\rangle =f=1/2.$ In addition, the variables $x_j,~u_j$ appearing in $h_i$ describe the STD and STF mechanisms, respectively. We assume that these variables evolve according to a well-known discrete dynamics~\cite{tsodyksNC,torresCAPACITY,mejias09}
\begin{equation}
x_j(t+1)=x_j(t)+\frac{1-x_j(t)}{\tau_{rec}}-U_{SE}u_j(t)x_j(t)s_j(t)
\label{x}
\end{equation}
\begin{equation}
u_j(t+1)=u_j(t)+\frac{1-u_j(t)}{\tau_{fac}}+(1-U_{SE}u_j(t))s_j(t),
\label{u}
\end{equation}
where the dynamics of $u_j(t)$ has been normalized with respect to \cite{mejias09} for simplicity. Here, $U_{SE}$ is a parameter related with the fraction of neurotransmitters released after the presynaptic neurons fires, and $\tau_{rec}, \tau_{fac}$ are, respectively, the time constants for depressing and facilitating mechanisms. The static-synapses situation (that is, when synapses do not display STD nor STF) may be obtained if one sets $x_i(t)=u_i(t)=1,~\forall~i,t$ or, alternatively, in the limit in which synaptic time constants $\tau_{rec},\tau_{fac}$ become too small (for more details on this limit, see \cite{mejias09}). The neuron firing thresholds are given by
\begin{equation}
\theta_i=\frac{1}{2} \sum_{j\neq i} \omega_{ij}.
\label{threshold}
\end{equation}
This last expression allows to recover the original Hopfield model when one considers static synapses (that is, the case $x_i(t)=u_i(t)=1,~\forall~i,t$).

\section{Results}

\begin{figure}[t!]
\psfig{file=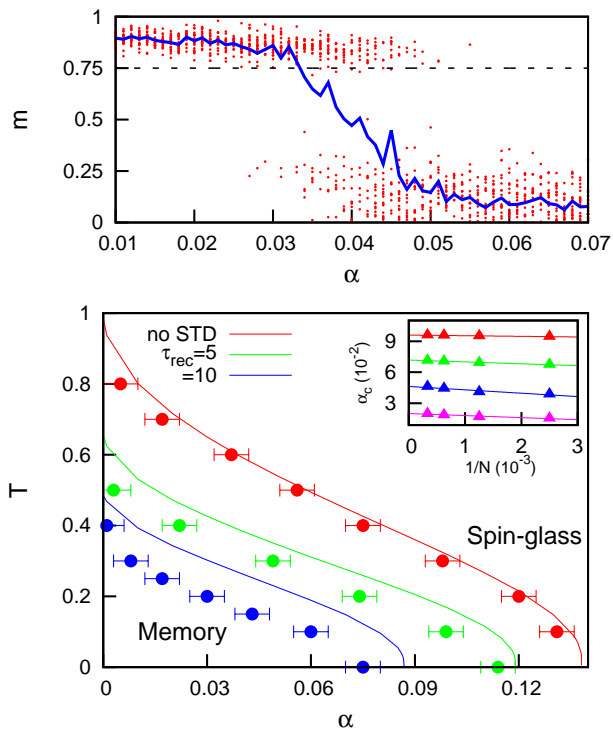,width=8cm}
\caption{Top: Criterion to compute the stability line between memory phase
and spin-glass phase in numerical simulations. Dots indicate individual
realizations, while the solid line indicates the macroscopic overlap
averaged over $20$ realizations. The dashed line in $m=0.75$ is
the threshold used as a criterion to discern between phases. Bottom:
Critical line, in the parameter space $\{T,~\alpha\}$, separating
the recall (memory) phase and the spin-glass phase for an ANN with
depressing synapses and $U_{SE}=0.1$. One may observe that the inclusion
of STD reduces the area of the memory phase. We show numerical simulations
(points) of an ANN of $N=3000$ neurons, and mean-field predictions
(lines). Inset: finite-size scaling analysis for some of the points
displayed in bottom panel (concretely, for $\tau_{rec}=5$). Lines
correspond, from top to bottom, to $T=0.1,~0.2,~0.3,~0.4$, respectively.
Network sizes were $N=400,~800,~1600$ and $3000$.}
\label{fig1}
\end{figure}
In order to numerically compute the phase diagrams of the system, one has to determine, for a given set of values $(T,\alpha)$, the steady-state values of the order parameters characterizing the behavior of the network. Following \cite{amitB,hertzB}, these order parameters are the overlap functions $m^{\mu} \equiv \frac{1}{N} \sum_j (2\xi_j^{\mu}-1) \langle 2 s_j -1\rangle $, the {\em spin-glass} order parameter $q \equiv \frac{1}{N} \sum_i \tanh^2 \left[ 2 \beta \left( h_i-\theta_i\right) \right]$, and the pattern interference parameter $r\equiv \frac{1}{\alpha}\sum_{\nu}^M (m^\nu)^2$, where the index $\nu$ denotes non-condensed patterns (i.e. satisfying $m^{\nu} \sim O(1/\sqrt{N})$). Typically, situations in which $m^{\mu} \sim O(1)$ for a given pattern $\mu$ (that is, the pattern $\mu$ has a {\em macroscopic} overlap), $m^{\nu} \sim O(1/\sqrt{N})$ for $\nu \neq \mu$, and $q \sim 1$, correspond to the recall phase. On the other hand, situations in which the state of the system has an overlap  $m^{\nu} \sim O(1/\sqrt{N})$ for all stored patterns and $q<1$ corresponds to the spin-glass phase \cite{amit87}. We are interested here in the critical line between memory and spin-glass phases, which delimits the area of the memory phase \cite{amitB}. A simple way to obtain these phase diagrams is to compute, for a given value of $T$, the maximal value of $\alpha$ for which the stationary value of the macroscopic overlap is $m \geq 0.75$. This criterion is usually taken for the numerical evaluation of storage capacities in neural networks \cite{amitB,amit87}, and is illustrated in top panel of Fig. \ref{fig1}. Equations corresponding to the mean-field solutions (see Appendix) may be solved by employing any standard minimization algorithm to drive the absolute value of the difference between both sides of each equation to zero.

Following these methods, we first compute the phase diagram of a stochastic neural network whose synapses present STD. In the following, together with numerical results, we show the predictions of a simplified mean-field approach previously introduced in \cite{mejias09}, to support our numerical results and provide some theoretical insight of the phenomenology (see Appendix for details). Fig. \ref{fig1} shows a diagram $\{ T,\alpha \}$ where we can see the critical line separating the memory phase and the spin-glass phase, for different values of the STD time constant and a fixed value of $U_{SE}$. The mechanism of STF is not present here (that is, $u_i(t)=1,~ \forall ~t,i$). As we can see, STD reduces the area of the memory phase with respect to the standard Hopfield model (that is, when STD is not present). Such reduction becomes more prominent as the depression time scale $\tau_{rec}$ increases, and indicates that STD, while being quite convenient for filtering and processing information in spike trains \cite{abbott97,tsodyks97} or for dynamic memories \cite{torresNC}, is not adequate for associative memory tasks. Previous works had already suggested this effect \cite{torresCAPACITY,mejias09}, although only for the particular case of deterministic ($T=0$) networks. Here, on the other hand, the adverse effect of STD on associative memory is shown for the more general and realistic case of stochastic ($T>0$) neural networks, a result which corroborates those of recent studies \cite{okada11}. We find that the reduction of the memory area is due to the temporary weakening of synaptic connections due to STD, which decreases the stability of the fixed points of the dynamics \cite{torresNC}.

\begin{figure}[t!]
\psfig{file=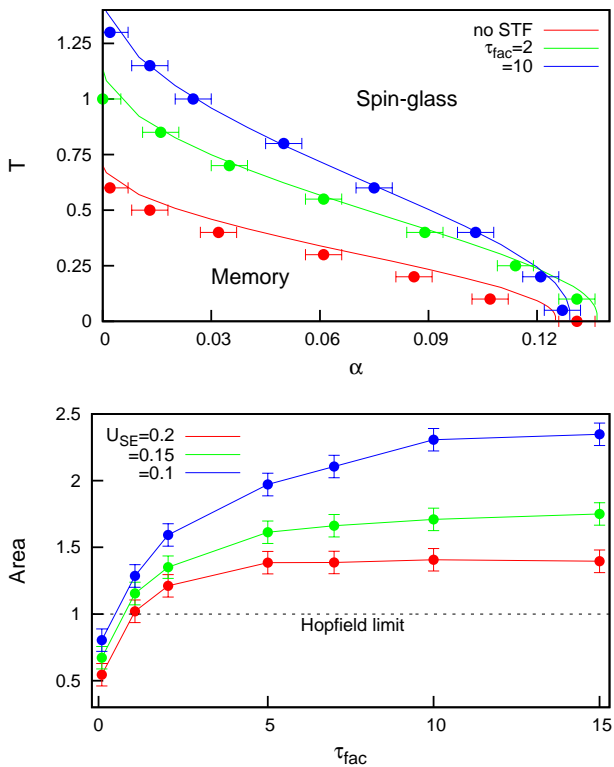,width=8cm}
\caption{
Top: Critical line $(T,~\alpha)$ separating the recall (memory) phase and the spin-glass phase for an ANN with dynamic synapses, concretely, for $\tau_{rec}=2$, $U_{SE}=0.2$ and several values of $\tau_{fac}$. One can observe that the inclusion of STF increases the area of the memory phase. Numerical simulations (points) of an ANN of $N=3000$ neurons are confirmed by mean-field predictions (lines). Bottom: Area of the memory phase as a function of $\tau_{fac}$, for $\tau_{rec}=2$ and different values of $U_{SE}$, obtained numerically by discretizing the memory phase diagrams from simulations. The area of the standard Hopfield model is indicated by a dashed line.}
\label{fig2}
\end{figure}

The effects of STF in the retrieval properties of stochastic neural networks may be investigated as well. Top panel of Fig. \ref{fig2} shows such effect, for a network with a fixed degree of STD ($\tau_{rec}=2$) and increasing levels of STF. As we can see, the net effect of STF is the enlargement of the memory phase as $\tau_{fac}$ increases, thus inducing an beneficial effect on associative memory tasks as opposed to STD. Such enlargement is even able to overcome the initial reduction of the memory phase caused by STD (red curve in top panel of the figure) and to expand the limits of the memory phase far beyond the critical line of the standard Hopfield model. This may be better appreciated in bottom panel of Fig. \ref{fig2}, where the area of the memory phase (namely, $A_m$) is computed as a function of $\tau_{fac}$ for different values of $U_{SE}$. For clarity purposes, we have normalized the area units so that the standard Hopfield model (that is, no short-term synaptic plasticity) has $A_m=1$. For any of the curves shown, one may see that, as $\tau_{fac}$ increases, the area also increases from $A_m < 1$ (the area is less than one due to the presence of STD), surpassing the Hopfield memory area at $\tau_{fac} \simeq 1$, and finally saturating at $A_m>1$ for $\tau_{fac} > 10$. This saturation limit on $A_m$ depends strongly on the synaptic parameters, and in particular this limit increases when the parameter $U_{SE}$ takes lower values, as the figure also shows. For very large values of $\tau_{fac}$, we have found that $A_m$ starts to slowly decrease, although such decay is only present when $\tau_{rec}>0$ (data not shown). If STD is not present, the effect of STF is a monotonic increment of $A_m$ with $\tau_{fac}$. The increment of memory area may be explained as follows: when the network is in (or close to) a previously stored pattern (or equivalently, in a fixed point of the dynamics), neurons which were active for such pattern start firing. Such firing induces, via STF dynamics, a temporal reinforcement of the synaptic connections of such neurons (as one may infer from Eq. (\ref{A3})), and this in turn improves the stability of such fixed point state. As a consequence, patterns which were unstable for networks with classical hebbian synapses are now stable in the presence of STF, and therefore the area of the memory phase is enlarged.

It is particularly interesting to note that facilitating synapses present typically low values of $U_{SE}$ compared with other synapses \cite{tsodyks97}, suggesting that they could be contributing to enhance the retrieval properties of the network. It is also worth noting that $A_m$ continues rising with $\tau_{fac}$ even for conditions in which STF no longer improves the storing abilities of deterministic networks, suggesting a nontrivial role of STF on the retrieval properties of the system. This may be seen for instance in top panel of Fig. \ref{fig2}, where the maximum storage capacity at $T=0$ is larger for $\tau_{fac}=2$ than for $\tau_{fac}=10$, although, as the bottom panel shows, the later has a larger value of $A_m$. We may conclude then that networks with facilitating synapses display larger memory phases than standard Hopfield networks, and are therefore able to efficiently retrieve previously stored information in more general conditions --and in particular, in the presence of strong noise.

\begin{figure}
\psfig{file=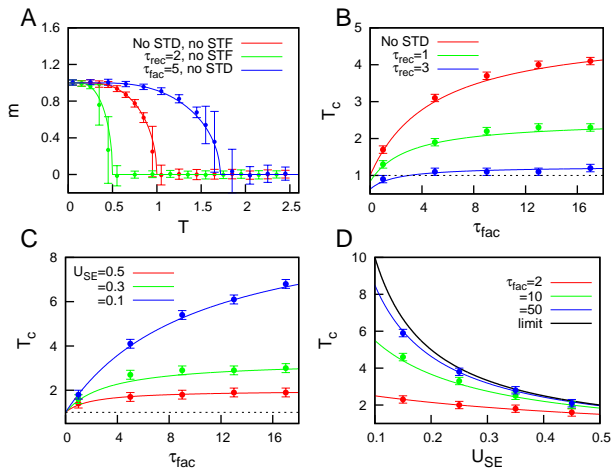,width=8cm}
\caption{(A) Steady state of the overlap function $m$ of an ANN, with dynamic synapses and one stored pattern ($\alpha \rightarrow 0$), as a function of $T$ for $U_{SE}=0.5$. (B) Critical temperature $T_c$ as a function of $\tau_{fac}$, for different values of $\tau_{rec}$ and $U_{SE}=0.5$. (C) Same as panel B, but with STF only (no STD) and different values of $U_{SE}$. (D) Dependency of $T_c$ with $U_{SE}$ for STF only (no STD) and different values of $\tau_{fac}$. Black line corresponds to the theoretical limit of large $\tau_{fac}$, where the critical temperature becomes $T_c = T_c^* = 1/U_{SE}$. In all panels, simulations of an ANN of $N=3000$ neurons (points) and mean-field predictions (lines) are displayed.}
\label{fig3}
\end{figure}

A more careful inspection of top panel in Fig. \ref{fig2} at low $\alpha$ values indicates that STF yields an enlargement of the memory phase towards high-temperature regions. In order to further explore this effect, we particularize to the case of $\alpha \rightarrow 0$. This limit corresponds to the case in which one considers a finite number of stored patterns $P$ (and therefore $\alpha=\frac{P}{N} \rightarrow 0$ in the thermodynamic limit $N \rightarrow \infty$). A relevant issue to consider here is the influence of dynamic synapses on the {\em critical temperature} $T_c$ for which the system passes from a recall (memory) phase to a non-recall phase. When $\alpha \rightarrow 0$, the non-recall phase corresponds to the {\em paramagnetic} phase, in which $m^{\mu}$ and $q$ are zero in the thermodynamic limit ($N \rightarrow \infty$) \cite{amit87}.  The critical temperature $T_c$ may be evaluated from a numerical point of view by computing the value of $T$ for which the steady state of the macroscopic overlap decays to zero \cite{amitB,perettoB}. Additionally, and to complement these numerical results, one may easily simplify our mean-field approach (see Appendix) by doing $\alpha \rightarrow 0$ and looking for the condition of nonzero solutions in Eqs. (\ref{m}-\ref{r}), which gives a critical temperature of the form
\begin{equation}
T_c=\frac{\gamma'}{1+\gamma \gamma'} \equiv \frac{1+\tau_{fac}}{1+U_{SE} (\tau_{rec}+ \tau_{fac}+ \tau_{rec} \tau_{fac})}.
\label{Tcrit}
\end{equation}
The evaluation of $T_c$ is depicted in Fig. \ref{fig3}A, which also shows the influence of the synaptic time constants $\tau_{rec}$ and $\tau_{fac}$ on the value of $T_c$. For the case of static synapses, the critical temperature takes the usual value $T_c=1$ \cite{amitB}. When STD is present, one observes that lower values of $T_c$ are obtained (around $T_c \simeq 0.5$ in the figure). Such result indicates that ANNs with depressing synapses require a low level of intrinsic stochasticity to retrieve stored information via associative memory processes, even for vanishing network load. On the other hand, the inclusion of STF leads to larger values of $T_c$ (around $T_c \sim 1.7$ in the figure), indicating that an optimal performance in retrieval tasks may be achieved by ANN with facilitating synapses even in the presence of high levels of intrinsic noise. This is also shown in panels B and C of Fig. \ref{fig3}, where one can see that $T_c$ is a monotonically increasing function of $\tau_{fac}$, for different values of $\tau_{rec}$ (panel B) or $U_{SE}$ (panel C). Again, we find that the drastic increment of the critical temperature is due to the temporal reinforcement of concrete synaptic connections via STF dynamics.

It is worth noting that the critical temperature of networks with STF comfortably surpass the critical temperature of the standard Hopfield network (namely, $T_c=1$), saturating at a certain value $T_c^*>1$ for large $\tau_{fac}$. Indeed, from Eq. (\ref{Tcrit}) we can obtain $T_c^*$ by considering $\tau_{rec} \rightarrow 0$ (which corresponds to absence of STD if one considers the limit of smooth dynamics of $x(t)$, see \cite{mejias09} for details) and $\tau_{fac} \rightarrow \infty$, obtaining $T_c^* =1/U_{SE}$. Such prediction is shown in Fig. \ref{fig3}D and confirmed with numerical simulations. This may be quite relevant since, as stated above, facilitating synapses in actual neural systems usually present very low $U_{SE}$ values \cite{tsodyks97}, which would correspond to high values of $T_c^*$. These findings indicate that dynamics synapses, and in particular STF, may play a key role in the retrieval of information in networks with low load and high noise conditions. 

\section{Discussion}

In this work we have shown the effect of short-term synaptic dynamics, and in particular STD and STF, in the recall properties of stochastic ANNs. Our findings indicates that STD has a negative effect on the retrieval abilities of the network, by decreasing the area of the memory phase. On the other hand, the presence of STF is able to enlarge this area far beyond the limits of the standard Hopfield model, highlighting a tremendous beneficial impact of STF on memory tasks. The effects of STF in the particular case of deterministic dynamics ($T=0$) have been addressed previously \cite{mejias09} and therefore they have not been discussed here, although our results are in agreement by them. On the other hand, the limit of vanishing network load ($\alpha \rightarrow 0$) has been studied in detail, revealing that the presence of STF yields arbitrarily high critical temperature values, with the maximum value given by $T_c^* = 1/U_{SE}$ for no STD and large enough $\tau_{fac}$. Such results indicate a potential role of STF in memory tasks in noisy situations. This is especially interesting when modeling working memory tasks, where neural populations have to maintain information actively in the form of activity patterns in high noisy conditions. In these situations, networks are known to handle simultaneously only a few activity patterns \cite{compte09}, a situation which could be modelled as ANN with low network load and high noise, as we have done here. 

Another surprising feature is that this increment of $T_c$ with $\tau_{fac}$ is simply due to the fast dynamics of STF, and does not involve having higher values of the absolute coupling strength. Indeed, the coupling strength of synapses is always upper-bounded in our model (that is, synaptic strengths can not be higher than those of the Hopfield model). The fact that such high performance in noisy conditions may be achieved by simply introducing a fast dynamics in the network links may have implications not only for neural network models, but for a wide class of cooperative networked systems with time-varying connections \cite{torres98,li10,kaufman98,tsekeris05,larhlimi11}.

Finally, it is worth noting that the good retrieval abilities of neural networks with facilitating synapses highlights a new way to develop efficient neural network paradigms. Indeed, good performance under high-noise conditions is a highly desirable feature from a technological point of view, and such improvement could be used to build more efficient Hopfield-based categorization systems, for instance.

\section{Appendix: Mean-field analysis}
\renewcommand{\theequation}{A.\arabic{equation}}
\setcounter{equation}{0}

In order to obtain an approximate mean-field theory, with the only purpose of orienting and supporting our numerical results, we follow a similar reasoning as the one presented in \cite{mejias09}. Briefly, from the definition of $h_i({\bf s},t)$ and Eqs. (\ref{hebb}) and (\ref{threshold}), we obtain
\begin{equation}
2[h_i({\bf s},t)-\theta_i]=\sum_{\mu} \epsilon_i^{\mu} \overline{m}^{\mu}({\bf s},t)-2 \alpha x_i(t) u_i(t) s_i(t) +\alpha
\label{ht}
\end{equation}
where $\epsilon_i^{\mu} \equiv 2\xi_i^{\mu}-1$, $\overline{m}^{\mu}({\bf s},t)\equiv\frac{1}{N}\sum_j \epsilon_j^{\mu} [2 x_j(t) u_j(t) s_j(t)-1]$ and, as stated above, $\alpha \equiv P/N$. Here, we consider only situations in which the system reaches a fixed point of the dynamics. A necessary condition for this is to keep relatively low values of $\tau_{rec}$, since when this time constant is large enough, the fixed point solutions of the dynamics (\ref{x}-\ref{u}) become unstable, leading to the appearance of {\em switching} behavior between activity patterns \cite{torresNC,bibitchkov02,cortes07}. On the other hand, when $\tau_{rec}$ takes relatively low values, the fixed points of the dynamics (\ref{x}-\ref{u}) are stable. In this situation, a plausible approximation is to completely neglect any temporal correlations in the dynamics (\ref{x}-\ref{u}). One may then consider both $x_i(t)$ and $u_i(t)$ as binary variables switching between two possible values, namely $\{x_+,u_+\}$ when $s_i(t)=1$, and $\{x_-,u_-\}$ when $s_i(t)=0$. If we choose such values to be the maximum and minimum values that $x_i(t)$ and $u_i(t)$ may reach, this will give us an estimation of their mean values as a function of the mean value of $s_i(t)$. Indeed, these maximum and minimum values may be easily computed by substituting $s_i(t)=1,~\forall t$, and $s_i(t)=0,~\forall t$, in Eqs. (\ref{x}-\ref{u}), yielding
\begin{equation}
\begin{array}{lll}
x_+=\frac{1}{1+\gamma \gamma'}, ~~~~x_-=1,
\\\\
u_+=\gamma', ~~~~u_-=1,
\end{array}
\end{equation}
where $\gamma\equiv U_{SE} \tau_{rec}$ and $\gamma'\equiv \frac{1+\tau_{fac}}{1+U_{SE}\tau_{fac}}$, as in \cite{matsumoto07,mejias09}. It is worth noting that, with this criterion, the values $\{x_+,~u_+ \}$ (which are associated with $s_i(t)=1,~\forall t$) correspond to the minimum values reached by $x_i(t),~u_i(t)$, and therefore $\{ x_-,~u_- \}$ correspond to the maximum values. Considering the above assumptions, one has
\begin{equation}
\begin{array}{lll}
x_i (t)\approx 1+\left( \frac{1}{1+\gamma \gamma '}-1 \right) s_i (t)
\\
u_i (t)\approx 1+(\gamma '-1) s_i (t).
\end{array}
\label{A3}
\end{equation}
We can therefore provide an estimation for the quantity $x_i(t)u_i(t)s_i(t)$ appearing in Eq. (\ref{ht}). Such estimation is given by
\begin{equation}
x_i(t)u_i(t)s_i(t) \approx \frac{\gamma '}{1+\gamma \gamma '} s_i (t) \quad \forall t.
\label{xFs}
\end{equation}
Note that this expression, once time-averaged, is equivalent to the one obtained for low temperatures \cite{mejias09} (although that was obtained under different conditions), and therefore the resulting mean-field will be valid for all possible values of $T$. 

Following a standard procedure (see \cite{mejias09} for further details), one may arrive to the approximate mean-field description of the network:

\begin{equation}
m=\left\langle \left\langle \tanh \left( \widehat{\beta} m + z \widehat{\beta} \sqrt{ \alpha r+\alpha \omega^2} \right) \right\rangle \right\rangle,
\label{m}
\end{equation}
\begin{equation}
q=\left\langle \left\langle  \tanh^2 \left( \widehat{\beta} m+ z \widehat{\beta} \sqrt{ \alpha r+\alpha \omega^2} \right) \right\rangle \right\rangle,
\label{q}
\end{equation}
\begin{equation}
r=\frac{q}{\left( 1-\widehat{\beta} (1-q)\right)^2}~,
\label{r}
\end{equation}
where $\left\langle \left\langle \cdots \right\rangle \right\rangle$ indicates an average over $z$ (the normal-distributed noise), variables $m,~q,~r$ are the order parameters of the model (see main text),  and the parameters $\widehat{\beta}$, $\omega$ are given, respectively, by $\widehat{\beta} \equiv \frac{\gamma'}{1+\gamma \gamma'} \beta$ and $\omega \equiv \frac{1+\gamma \gamma'-\gamma'}{\gamma'}$. As we have already stated, Eqs. (\ref{m}-\ref{r}) constitute a strongly simplified mean-field description of the model, used to support our numerical findings. Therefore, it should not be considered as a general mean-field solution of ANNs with dynamic synapses, due to the existence of switching behavior \cite{torresNC,cortes07} which cannot be explained within this framework. Extensions of this theory to account for switching dynamics constitutes an interesting research line that remains open.

\acknowledgments
We acknowledge financial support from Spanish MICINN project FIS2009-08451, MICINN CEI-GREIB translational project GREIB.PT$\_$2011$\_$19, Junta de Andaluc\'{\i}a project P06–FQM–01505, and NSERC Canada Discovery Accelerator Supplement Program.


\end{document}